\documentstyle[12pt,epsfig,amsfonts,latexsym]{article}

\makeatletter

\@addtoreset{equation}{section} \makeatother

\newcommand{\be}{\begin{equation}}

\newcommand{\ee}{\end{equation}}

\newcommand{\ba}{\begin{eqnarray}}

\newcommand{\ea}{\end{eqnarray}}

\newcommand{\bea}{\begin{eqnarray*}}

\newcommand{\eea}{\end{eqnarray*}}

\newcommand{\ti}{\tilde}

\newcommand{\gsim}{\raise.3ex\hbox{$>$\kern-.75em\lower1ex\hbox{$\sim$}}}

\newcommand{\lsim}{\raise.3ex\hbox{$<$\kern-.75em\lower1ex\hbox{$\sim$}}}

\begin{document}

\titlepage

\begin{flushright}
\today \\
hep-th/yymmnnn \\
\end{flushright}
\vskip 1cm

\begin{center}
{\Large \bf Spherical Collapse in Chameleon Models}
\end{center}
\vskip 1cm
\begin{center}
Ph.~Brax$^{a}$\footnote{{\tt brax@spht.saclay.cea.fr}}, R.~Rosenfeld$^{b}$\footnote{{\tt rosenfel@ift.unesp.br}}
and D.~A.~Steer$^{c}$\footnote{{\tt daniele.steer@apc.univ-paris7.fr}}\\
\vskip 5pt \vskip 3pt
{\it a}) {\em
Institut de Physique Th\'eorique, CEA, IPhT, CNRS, URA 2306,
  F-91191Gif/Yvette Cedex, France}
\vskip 3pt
{\it b}) {\em Instituto de F\'{i}sica Te\'orica, Universidade Estadual Paulista, Rua Dr. Bento T. Ferraz, 271, 01140-070, S\~ao Paulo, Brazil}
\vskip 3pt
{\it c}) {\em APC (UMR 7164, CNRS, Universit\'e Paris 7), 10 rue Alice Domon et L\'eonie Duquet, 75205 Paris Cedex 13, France}
\end{center}
\vskip 2.0cm

\begin{center}
{\bf Abstract}
\end{center}
We study the gravitational collapse of an overdensity of nonrelativistic matter under the
action of gravity and a chameleon scalar field. We show that the spherical collapse model is modified by
the presence of a chameleon field.
In particular, we find that even though the chameleon effects can be potentially large at small scales,
for a large enough initial size of the inhomogeneity
the collapsing region possesses a thin shell that shields the modification of gravity induced by the
chameleon field, recovering the standard gravity results. We analyse
the behaviour of a collapsing
shell in a cosmological setting in the presence of a thin shell and find that, in contrast to the usual case, the critical density  for
collapse in principle depends on the initial comoving size of the inhomogeneity.

\newpage
\section{Introduction}

Fundamental scalar particles have not been found in Nature yet. Many examples  have been postulated over the last forty years; the most illustrious  being the Higgs boson, a cornerstone of the Standard Model (SM),
which may well be the first one to be discovered   at the Large Hadron Collider. The existence of scalar fields has  also been invoked  in order to explain
the inflationary phase in the early universe\cite{lyth}. In the same vein, the late time acceleration of the  expansion of the universe could result from the presence of dark energy scalar fields\cite{Copeland:2006wr,caldwell,trodden,Brax:2009ae}. In all these cases, the coupling of  scalar particles to normal matter
leads to new interactions and possibly variations of parameters such as masses and coupling constants \cite{uzan}.

The physics of scalar fields coupled to matter depends crucially on the scalar field mass. For instance, the Higgs boson is expected to have a large mass ($m_h > 115$ GeV) compared to the Hubble scale today
($H_0 = 10^{-42}$ GeV) implying  that the new interaction mediated by the Higgs boson is of very short range. Moreover  the Higgs field has
been frozen at the minimum of its potential since the early stages of the universe. On the other hand,
several extensions of the Standard Model  predict the existence of light
scalar fields. Some of these field might be moduli of string compactifications or the dilaton.
Another class comprises  the  fields which can appear as a consequence of
the spontaneous symmetry breaking of a global quasi-exact symmetry.
These pseudo-Goldstone bosons may play a relevant role in explaining the recent stage of acceleration
observed in cosmological data. In this case, the scalar fields are nearly massless as their mass should be of the order of the
Hubble rate now.
All these light scalar fields can couple to matter  and hence introduce new long range forces
which would appear as a modification of gravity. The new interaction would also induce a non-conservation of
the matter field stress-energy tensor, which can be interpreted as arising from  mass-varying matter fields.
Limits on this type of interactions arise from SNIa and CMB data \cite{DEInt}.
As we will see below, alternatively one can choose to work in the so-called Jordan frame, where the
matter field stress-energy tensor is conserved. In this case, Newton's constant varies and its variation since Big Bang Nucleosynthesis
should not exceed some forty percent \cite{uzan}.

Modifications of gravity should not be taken lightly.
There are stringent limits on the couplings of scalar fields to
normal baryonic matter arising from the search of a fifth force or deviations from the usual
gravitational force at different scales, from the laboratory to the solar system.  In particular, the Cassini bound~\cite{Bertotti:2003rm}
implies that the coupling to matter should be less than $10^{-5}$.
Several mechanisms have been proposed to alleviate this problem.
The first one which applies to the dilaton in the strong string coupling limit is
due to Damour and Polyakov \cite{DP} and it asserts that the field is attracted by the zero coupling limit cosmologically.
Another mechanism occurs in the Dvali-Gabadadze-Porrati model \cite{DGP}
whereby the brane bending mode, whose presence would lead to strong deviations from Newtonian gravity, is suppressed at short distance thanks to the
Vainshtein mechanism.
Another possibility  is to generate a mass term for the scalar field
which  depends on the matter density of the environment. This is implemented in the so-called chameleon
mechanism \cite{Khoury:2003rn,braxkhoury}, where a coupling between a scalar field (the chameleon field) and matter
arises from a
conformal rescaling of the metric in the matter sector.
The nonlinear interaction of the chameleon field generates a shielding, known as the thin-shell effect,
that diminishes the modifications of gravity for dense and large enough bodies.

In this paper we are interested in modifications of General Relativity in chameleon models, focusing
on the possible consequences for the growth and collapse of large scale structures in the universe.
The growth of perturbations in chameleon models in the linear regime was studied in \cite{BraxLinear}.
Recently, there has been a new bout of  interest in studying the influence of modified gravity models on the
formation of large scale structures. Indeed most of the aforementioned models of modified gravity show very little deviation from a pure $\Lambda$CDM model since Big Bang Nucleosynthesis. Differences only emerge at the perturbative level in the way structures grow. In that respect it is of high interest to go beyond the
linear level and study the nonlinear regime of structure formation. A semi-analytical tool that has
been used to analyse the nonlinear stages of gravitational collapse
is the so-called spherical collapse model \cite{SphColl}.

The spherical collapse model has been recently used in
models with a simple
Yukawa-type modification of gravity \cite{MartinoStabenauSheth}, in the so-called $f(R)$ models
 \cite{SchmidtLimaOyaizuHuIII}, in braneworld cosmologies \cite{SchmidtLimaHu} and in models which
allow for dark energy fluctuations \cite{DEFluc}.
We will focus on the collapse of objects with a thin shell that arises in the context of chameleon models.

This paper is organized as follows. In section 2 we give a brief introduction to chameleon models and  discuss the modifications of gravity which occur in these
models, both for geodesics and the growth of perturbations. In section 3 we solve the dynamical equation for the scalar field in a spherical configuration of matter overdensity, finding its profile and studying the
effects of the existence of a thin shell which effectively shields the modifications of gravity. These profiles serve as templates for the chameleon behaviour during the spherical collapse.
Section 4 presents an analytical discussion of the consequences of the chameleon models for
the spherical collapse models of large scale structure formation. Phenomenological consequences are expounded in section 5 with a
simple example before presenting our conclusions.

\section{Modification of Gravity with Chameleons}

\subsection{Chameleon models}

The action governing the dynamics of the chameleon field $\phi$  is given by
\be
S=\int d^4x\sqrt{-g}\left\{\frac{M_{\rm Pl}^2}{2}{\cal
R}-\frac{1}{2}g^{\mu \nu} \partial_\mu \phi \partial_\nu \phi- V(\phi)\right\} +
 \int d^4x \sqrt{-\tilde{g}} {\cal
L}_m(\psi_m^{(i)},\tilde g_{\mu\nu})\,, \label{action}
\ee
where $g_{\mu \nu}$ is the metric in the Einstein frame, $\tilde g_{\mu\nu}$ is the metric in the Jordan frame,
$M_{\rm Pl}\equiv (8\pi G_N)^{-1/2}\equiv {\kappa_4^{-1}}$ is the reduced Planck mass, ${\cal R}$ is
the Ricci scalar for the Einstein-frame metric, and $\psi_m^{(i)}$ are various matter fields
labeled by $i$.
The Einstein-frame metric $g_{\mu \nu}$ and Jordan-frame metric $\tilde g_{\mu\nu}$ are related
by the conformal rescaling
\be
\tilde g_{\mu\nu}=e^{2\beta\phi/M_{\rm Pl}}g_{\mu\nu}\, \equiv \Omega^2(\phi) g_{\mu\nu},
\label{conformal}
\ee
where $\beta$ is a  dimensionless constant~\cite{dgg}. In chameleon models this constant
is assumed to be the same for all types of matter, respecting the weak equivalence principle where
all material particles follow the same metric.
This conformal coupling of $\phi$ with matter particles is a key ingredient of the model, and as a result the
excitations of each matter field $\psi_m^{(i)}$ follow the
geodesics of the Jordan frame metric.
As a consequence, the physical energy momentum tensor of matter $\tilde{T}^{(m)}_{\mu \nu}$ is defined in the Jordan frame \cite{Damour1,Damour2}, where it is covariantly conserved
\be
\tilde{\nabla}_\mu \tilde{T}^{(m)\mu}_{\; \; \; \; \; \; \; \;\nu} = 0.
\label{consvJ}
\ee
It is related to the energy-momentum tensor in the Einstein frame, $T^{(m)}_{\mu \nu}$, by
$
\tilde{T}^{(m)\mu}_{\; \; \; \; \; \; \; \;\nu} =\Omega^{-4}{T}^{(m)\mu}_{\; \; \; \; \; \; \; \;\nu} .
\label{EtoJ}
$

In this paper we restrict our attention to {\it non-relativistic matter} for which the only non-vanishing component of $ \tilde{T}^{(m)\mu}_{\; \; \; \; \; \; \; \;\nu} $ is $ \tilde{T}^{(m)0}_{\; \; \; \; \; \; \; \;0} =-\tilde{\rho}$, giving an energy density of matter $\rho =\Omega^4 \tilde{\rho}$ in the Einstein frame.
In the Einstein frame, the equations of motion following from Eq.~(\ref{action}) are then
\ba
G_{\mu \nu} &=& \frac{1}{  M_{\rm Pl}^2} \left[
{T}^{(m)}_{\mu \nu}
+ \partial_\mu \phi \partial_\nu \phi - \frac{1}{2} \partial^\alpha \phi \partial_\alpha \phi g_{\mu \nu}- V g_{\mu \nu} \right]
\label{einstein}
\\
\nabla_\mu \nabla^\mu \phi &=&
 \frac{dV}{d\phi} - \frac{\Omega'}{\Omega} {T}^{(m)}.
\label{KGeqn}
\ea
where ${T}^{(m)}=g^{\mu \nu} {T}^{(m)}_{\mu \nu}  = -\ti\rho \, \Omega^{4} $.
Now consider a spatially flat Friedmann-Lemaitre-Robertson-Walker (FLRW) geometry with Einstein frame metric
\be
ds^2 = -dt^2 + a(t)^2 dx_i dx^i,
\ee
where $t$ is cosmic time.   Then, for nonrelativistic matter and a spatially uniform scalar field, Eqs.~(\ref{einstein}), (\ref{KGeqn}) and (\ref{consvJ}) reduce to
\ba
H^2 \equiv \left( \frac{\dot{a}}{a} \right)^2 &=&  \frac{1}{ 3 M_{\rm Pl}^2} \left[ \Omega \hat{\rho} + \rho_\phi \right]
\label{Fried}
\\
\frac{\ddot{a}}{a}& = &-\frac{1}{ 6 M_{\rm Pl}^2} \left[ \Omega \hat{\rho} + \rho_\phi + 3  p_\phi \right]
\\
\ddot{\phi} + 3 H \dot{\phi}
&=& -\frac{dV}{d\phi} - \Omega'  \hat{\rho}
\label{KGeqn2}
\\
\dot{\hat \rho} + 3H \hat \rho&=&0.
\label{consvhat}
\ea
Here $\Omega' = d\Omega/d\phi$, a dot denotes a derivative with respect to cosmic time $t$, and the energy density of the chameleon field is $\rho_\phi ={\dot{\phi}^2}/{2} + V$ whilst its pressure is $p_\phi={\dot{\phi}^2}/{2} - V$. The quantity $\hat{\rho}$ is defined by
\be
\hat{\rho}\equiv \Omega^{-1} \rho
\ee
and is conserved as a consequence of (\ref{consvJ}) (for non-relativistic matter only). Equivalently the Einstein frame energy density $\rho$ satisfies
\be
\dot \rho+ 3H \rho= \beta\kappa_4 \rho\dot \phi .
\label{rhoEeq}
\ee

It is important to draw a distinction between the three energy densities $\rho$, $\hat \rho$ and $\tilde \rho$. The latter is the Jordan
frame energy momentum which is always conserved, independently of the equation of state of matter.
As a result, the density $\tilde \rho$ will appear in the equations for conservation of matter
and for the fluid velocity. The Einstein density $\rho$ is the source of the Newton potential as
it appears in the Poisson equation. For non-relativistic matter, the chameleon field dynamics are influenced
by the conserved matter density $\hat\rho$. It appears as a source term which modifies the effective potential of the chameleon.

Indeed from the Klein-Gordon  equation one can define an effective potential for the chameleon,\footnote{Here we assume that, since $\hat{\rho}$ is conserved, $d\hat{\rho}/d\phi=0$.}
\be
V_{\rm eff}(\phi) = V(\phi) +  \hat {\rho}\Omega =   V(\phi) +  \hat {\rho} e^{\beta\phi/M_{\rm Pl}}.
\label{veff}
\ee
It follows that $V_{\rm eff}$ has a minimum at $\phi_{min}(\hat{\rho})$ which is solution to
\be
\frac{dV}{d\phi}= -\frac{\beta}{M_{\rm Pl}} {\rho}.
\label{minphi}
\ee
Notice that the source term is the Einstein energy density which depends on the chameleon field via the conformal factor $\Omega$, since $\rho = \Omega \hat{\rho}$.
The mass at the minimum is given by
\be
m^2= \frac{d^2 V}{d\phi^2}+ \frac{\beta^2}{M_{\rm Pl}^2} {\rho}.
\label{msqdef}
\ee
As $\frac{dV}{d\phi}<0$ and $\frac{d^2V}{d\phi^2}>0$ for successful chameleon models \cite{Khoury:2003rn},
\be
m^2\ge 3 \beta^2 H^2
\ee
where we have assumed that the chameleon energy density is negligible compared to ${\rho}$, $\rho_\phi \ll \rho$.

In contrast to usual quintessence models, in most cases the mass of the chameleon
is much larger than the Hubble parameter. An explicit example will be given in section 5.
Thus it is possible to have fluctuations in the chameleon field at scales much smaller than
the horizon scale which can be of cosmological importance for structure formation.

The minimum of the effective potential is an attractor,
and the chameleon therefore settles down there, and evolves slowly since at least Big Bang Nucleosynthesis.
In that case, the potential energy of the chameleon is nearly constant and the model is equivalent to a $\Lambda$CDM model. We will see that the dynamics of chameleon models are not equivalent to
a $\Lambda$CDM model at the level of perturbations and when the overdensities experience a spherical collapse phase.

\subsection{Geodesics}
We are interested in the dynamics of  CDM in the presence of a chameleon field background.
The trajectories of the CDM  particles satisfy the geodesic equation in the Jordan frame
\be
\frac{d^2x^\mu}{d{\ti\tau}^2}
+\tilde{\Gamma}^{\mu}_{\nu \rho}\frac{dx^\nu}{d{\ti\tau}}\frac{dx^\rho}{d{\ti\tau}} =0
\label{geoJ}
\ee
where
\be
d{\ti \tau}^2 = - \tilde{g}_{\mu \nu}dx^\mu dx^\nu
\ee
and the Christoffel symbol are
calculated with the metric $\ti g_{\mu\nu}$.
In the following we
consider the perturbed Jordan frame metric (in conformal-Newtonian gauge)
\be
d\ti{s}^2= \tilde{a}^2(\eta) \left[ -(1+2\phi_N) d\eta^2 + (1-2\phi_N)dx^idx_i \right],
\label{pertJ}
\ee
where $\eta$ is the conformal time and the Jordan frame scale factor $\tilde{a}$ is related to the Einstein frame scale factor $a$ by
\be
\tilde{a} = \Omega a.
\ee
The scalar metric perturbation $\phi_N$ is
related to the perturbation in the Newtonian gravitational potential and is sourced by the CDM density perturbation $\delta$ defined through
\be
\tilde \rho=\tilde  \rho_\infty (1+\delta)
\label{pert}
\ee
where $\tilde{\rho}_\infty$ is the background energy density.
Notice that when $\phi\ll M_{\rm Pl}$, the Jordan frame metric can be expanded as
\be
d\ti{s}^2= {a}^2(\eta) \left[ -(1+2\phi_N+2\beta \kappa_4 \phi) d\eta^2 + (1-2\phi_N+ 2 \beta \kappa_4 \phi)dx^idx_i \right].
\label{pertJ}
\ee
Hence, in the Jordan frame in which particles couple to gravity,  two distinct Newtonian potentials appear whose difference comes from the
chameleonic field. We will see that the existence of a chameleonic contribution to Newton's potential identified from the $d\eta^2$ element affects the motion of particles along geodesics.

Before continuing, we now specify certain approximations which will be made throughout the remainder of this paper. First, we assume that
\be
\phi_N\ll 1
\label{pertN}
\ee
 during structure formation. Then we assume that:
\be
\phi'_N \ll \tilde{\cal H}, \qquad \frac{\phi'}{M_{\rm Pl}}\ll \ti{\cal H}
\label{jap}
\ee
where $'=\partial/\partial \eta$ and $\ti{\cal H} ={d\ln \ti{a}}/{d\eta}$.  This implies, for example, that the Hubble parameters in the Einstein and Jordan frame are essentially identical:
$\tilde{\cal H} = {\cal H} +{\beta \phi'}/{M_{\rm Pl}} \simeq {\cal H}$. It also implies from (\ref{rhoEeq}) that $\rho$, as well as $\hat{\rho}$, are approximately conserved.  This is compatible with our next approximation, namely that
\be
\frac{\phi}{M_{\rm Pl}} \ll 1 \qquad \Longrightarrow \qquad \Omega(\phi) \simeq 1.
\label{dd}
\ee
Finally, in the non-relativistic limit, all spatial gradients dominate over time variations, so that for example
\be
|\partial_i \phi| \gg \phi'.
\ee

Using (\ref{pertN}) and (\ref{jap}), the geodesic equation in the Jordan frame, Eq.~(\ref{geoJ}),
can be simplified. From (\ref{pertJ}) the Christoffel symbols are given by
\be
\tilde \Gamma^{i}_{00}= \tilde{a}^2 \partial^i\Psi, \qquad \tilde \Gamma^i_{j0}\approx \tilde{\cal H} \delta^i_j
\ee
where
\be
\Psi= \phi_N + \beta \kappa_4\phi.
\label{gravp}
\ee
Hence the geodesic equation becomes
\be
\frac{d^2 x^i}{d\tilde{\tau}^2}+ 2\frac{\ti{\cal H}}{\ti{a}} \frac{dx^i}{d{\tilde{\tau}}} = -\partial^i \Psi.
\label{geoo}
\ee
where we have used that
\be
dx^0 \equiv d\eta = \frac{1}{\ti{a}} d\ti{\tau}
\ee
from the zero component of the geodesic equation.
Thus in proper time and in comoving coordinates, particles follow Newtonian trajectories under the influence of the effective  gravitational potential $\Psi$ given in Eq.~(\ref{gravp}).
In other words, matter moves under the influence of a new Newtonian potential modified
by the profile of the scalar field.

\subsection{Perturbations}
Chameleon theories are very different from a cosmological constant
at the perturbative level. This can have important effects
on structure formation, as we will see shortly.

The Lagrangian description of trajectories is equivalent to an Eulerian one in which the velocity field $v^i= \frac{\partial x^i}{\partial\eta}$ obeys the Euler equation. Let us define the unit 4-velocity
\be
u^\mu=\frac{dx^\mu}{d\ti{\tau}} \qquad \qquad (u^\mu u^\nu \tilde{g}_{\mu \nu} = -1)
\ee
so that in the non-relativistic limit, $u^i = v^i/\tilde{a}$.
Then using the chain rule, we obtain
\be
\frac{d^2x^i}{d\ti{\tau}^2} =   \frac{1}{\ti{a}^2}\left [{v^{i\prime }}- \ti{\cal H}v^i + v^j\partial_j v^i\right ].
\ee
Thus the geodesic equation reduces to the Euler equation
\be
{v^{i \prime}} + v^j\partial_j v^i= -\ti{\cal H} v^i -\delta^{ij}\partial_j \Psi
\label{Euler}
\ee
for the nonrelativistic fluid. Notice the presence of the effective Newton potential $\Psi$.

The divergence of (\ref{Euler}) gives
\be
\Theta' + \ti{\cal H} \Theta + \frac{\Theta^2}{3} = -\Delta \Psi
\label{theta}
\ee
where $\Theta = \partial_i  v^i$. Furthermore, energy conservation in the Jordan frame implies
that the density perturbation $\delta$, defined in Eq.~(\ref{pert})
satisfies
\be
\delta' + \Theta (1+ \delta) = 0.
\label{delta}
\ee
Combining eq.(\ref{theta}) and eq.(\ref{delta}) leads to a nonlinear equation for
the evolution of perturbations:
\be
\delta'' +  \ti{\cal H} \delta' - \frac{4}{3} \frac{\delta'^2}{1+\delta} = (1+\delta) \Delta \Psi.
\label{nonlineardelta}
\ee
We will use this equation in order to study the spherical collapse.

\subsection{Modified gravity}

For the following discussion it will be
more transparent  to
use physical coordinates
\be
r^i=\ti{a} x^i \qquad {\rm and} \qquad \nabla_i= \partial/\partial r^i.
\ee
Then the geodesic equation Eq.~(\ref{geoo}) is rewritten as
\ba
\frac{d^2 r^i}{d\ti\tau^2} &=& -\nabla^i \Psi + \frac{1}{\ti{a}} \frac{d^2\tilde{a}}{d\ti\tau^2}r^i
\nonumber
\\
&=& -\nabla^i \left( \Psi -  \frac{1}{2\ti{a}} \frac{d^2 \tilde{a}}{d\ti\tau^2}r^2 \right)
\nonumber
\\
 &\equiv& -\nabla^i \Psi_N.
\label{geo}
\ea
This is just Newton's law in an expanding universe, and matter moves under the
influence of the modified Newtonian potential $\Psi_N$. On using  Eq.(\ref{gravp}),
\ba
\Psi_N &=& \left( \phi_N - \frac{1}{2\ti{a}} \frac{d^2\tilde{a}}{d\ti\tau^2}r^2  \right) + \beta \kappa_4 \phi
\nonumber
\\
&\equiv &\Phi_N +  \beta \kappa_4 \phi.
\label{defPhi}
\ea

The `total Newtonian' potential $\Phi_N$ satisfies the Poisson equation sourced by the total density of matter. Indeed, let us now work under the assumptions given in Eqs.~(\ref{pertN})-(\ref{dd}).  Then from the zero-zero component of Einstein's equations (\ref{einstein}) in the Einstein frame, $\phi_N$ is solution to
\be
\Delta \phi_N = \frac{\kappa_4^2}{2} \rho_\infty \delta,
\ee
whereas we also have
\be
\frac{1}{\ti{a}} \frac{d^2\tilde{a}}{d\ti \tau^2} \approx  \frac{\ddot a}{a}=-\frac{4\pi G_N}{3} (\rho_\infty +(1+3w_\phi) \rho_\phi).
\ee
Thus from Eq.~(\ref{defPhi}) it follows the Poisson equation reads
\be
\Delta \Phi_N =\frac{\kappa_4^2}{2} \left[\rho +(1+3w_\phi)\rho_\phi \right]  \approx \frac{\kappa_4^2}{2} \rho
\label{totalPhi}
\ee
where we can neglect the chameleon energy density until very recently in the history of universe.

The resulting modifications of gravity are drastically illustrated for
models with $V=0$ in Minkowski space and $\rho=m_0\delta^{(3)}$, namely a massive point particle.
In this case the  Klein-Gordon equation (\ref{KGeqn}) reduces to
\be
\Delta \phi= \beta \kappa_4 m_0\delta^{(3)} \qquad \Longrightarrow \qquad \phi= -\frac{\beta \kappa_4 m_0}{4\pi r}
\ee
Thus from Eq.~(\ref{gravp}),
\be
\Psi= - \left(1+2\beta^2\right)\frac{G_N}{r} m_0.
\ee
Hence there is a modification of Newton's law depending on the coupling $\beta$
\be
G_{N,{\rm eff}}=\gamma G_N,\
\ee
where
\be
\gamma-1=2\beta^2
\ee

Of course, large $\beta$'s would be excluded by the Cassini bound
on the existence of fifth forces. This is not  the case for chameleon theories. Indeed two mechanisms are at play. The first
is that the chameleon develops a density dependent minimum, and at the minimum its mass increases with the density of the environment. This is sufficient in the atmosphere where the density is large, and this implies that there
is no effect on Galileo's Pisa tower experiment.
In a sparse environment such as the solar system or a laboratory vacuum chamber, a second mechanism must be
invoked: the existence of a thin shell which implies that $\gamma -1 \ll 2\beta^2$ for massive bodies
such as the moon, the sun and laboratory test bodies.

\section{Chameleon Profiles in Spherical Matter Configurations}
\label{sec:prof}

\begin{figure}
\centerline{\includegraphics[width=0.42\textwidth]{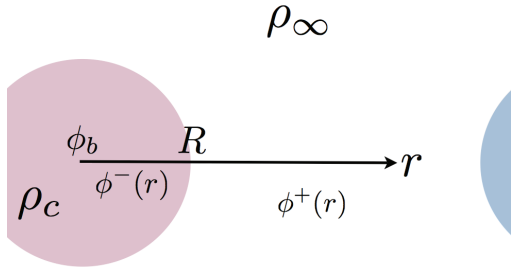}}
\caption{Scalar field profile with no thin-shell in the presence of a spherical overdensity with constant
density $\rho_c$ and radius $R$
immersed in a background of density $\rho_\infty$.}
\label{fig:1}
\end{figure}

We now find approximate solutions of the Klein-Gordon equation Eq.~(\ref{KGeqn}) in physical coordinates.  More precisely, in this section we assume that the matter source is a
{\it perfect sphere of constant radius $R$ with constant homogeneous density $\rho_c$}.  It is immersed in a homogeneous cosmological background of density $\rho_\infty \ll \rho_c$(see figure \ref{fig:1}). On sub-horizon scales where one can neglect time derivatives in the
evolution equation of the chameleon, one can treat this case as a quasi-static problem.  Our analysis is a generalisation of that presented in \cite{Khoury:2003rn}.

For the following discussion it will be useful to define the field values $\phi_c$ and $\phi_\infty$ as those which extremise the chameleon effective potential for the densities $\rho_c$ and $\rho_\infty$ respectively. Hence from  Eq.~(\ref{minphi}),
\be
V'_{c,\infty}\equiv \left. \frac{d V}{d\phi}\right|_{\phi_{c},\phi_\infty} = -\beta \kappa_4 \rho_{c,\infty}.
\ee
%Notice that generally $\phi_c < \phi_\infty$ and $m_c^2 > m_\infty^2$.

\subsection{Small bodies, no shell}

First we assume that the sphere is small (in a sense which will be made precise below) and its Newton potential at the surface is small also. Let $\phi_b$ denote the value of the field at the core.  We expect $\phi_b$ to differ significantly from $\phi_c$ and indeed that it should be closer to $\phi_\infty$ in the cosmological background.

Inside the overdensity, the Klein-Gordon equation Eq.~(\ref{KGeqn}) reduces to
\be
\frac{d^2\phi^-}{dr^2}+ \frac{2}{r} \frac{d\phi^-}{dr}= \frac{dV}{d\phi^-} + \beta \kappa_4 \rho_c.
\label{d1}
\ee
We expect
\be
\phi^-(r) = \phi_b + \delta \phi^-(r)  \qquad {\rm with} \qquad \delta \phi^-(r) \ll \phi_b
\ee
where  the value of the field at the core, $\phi_b=\phi^-(0)$, must be determined.  Substitution into (\ref{d1}) and expanding to first order gives
\be
\frac{d^2\phi^-}{dr^2}+ \frac{2}{r} \frac{d\phi^-}{dr}=m^2_b(\phi^- -\phi_b) + V'_b-V'_c
\ee
where
$V'_{b}= \partial_\phi V(\phi_{b})$ and $m_b^2 \equiv V''(\phi_b)$.
Thus the solution inside is simply
\be
\phi^-=\phi_b +\frac{V'_c-V'_b}{m^2_b} \left( 1- \frac{\sinh m_b r}{m_br}\right)
\label{sss}
\ee
where the two integration constants have been fixed through the conditions $d\phi^-/dr=0$ at $r=0$ and $\phi^-(0)=\phi_b$.

Outside the overdensity, $\phi^+=\phi_\infty + \delta \phi$, and following the same procedure leads to the equation of motion
\be
\frac{d^2\phi^+}{dr^2}+ \frac{2}{r} \frac{d\phi^+}{dr}=m_\infty^2(\phi^+-\phi_\infty)
\ee
where $m_\infty^2 \equiv V''(\phi_\infty)$, with solution
\be
\phi^+=\phi_\infty +\frac{De^{-m_\infty (r-R)}}{r}.
\label{outside}
\ee
Here $D$ is another integration constant and we have imposed that $\phi^+(r\rightarrow \infty)=\phi_\infty$.

The two solutions must match on the surface of the overdensity, namely
\be
\phi^+(R)=\phi^-(R) \qquad \left. \frac{d \phi^+}{dr}\right|_{R} = \left. \frac{d \phi^-}{dr}\right|_{R} \, .
\ee
This leads to
\be
D=\frac{V'_b-V'_c}{m^3_b(1+m_\infty R)}(\sinh m_b R- m_b R \cosh m_b R)
\end{equation}
and
\begin{equation}
\phi_b= \phi_\infty +\frac{V'_b-V'_c}{m^2_b}\left (1-\frac{\sinh m_b R}{m_b R}+ \frac{1}{(1+m_\infty R)}\frac{\sinh m_b R- m_b R \cosh m_b R}{m_bR}\right )
\end{equation}
In most interesting applications, the overdensity is small compared to the Compton wavelength $m_\infty^{-1}$.

On assuming that both $m_\infty R\ll 1$ and $m_b R\ll 1$, these give
\ba
D &=& \frac{R^3}{3} (V'_c-V'_b)
\label{D}
\\
\phi_b &=& \phi_\infty + \frac{1}{6} (V'_c-V'_b)R^2.
\label{phib}
\ea
This second equation determines $\phi_b$: on searching for a solution of the form
$\phi_b = \phi_\infty +\delta\phi_\infty $ where $\delta\phi_\infty  \ll \phi_\infty$ we find
\be
\phi_b = \phi_\infty  -  \frac{\beta}{3 M_{\rm pl}}  (\Delta \rho) R^2 \qquad (\Delta \rho = \rho_c - \rho_\infty ).
\ee
Using the condition $\phi_c\le \phi_b$ we find that this description is valid provided
\begin{equation}
\kappa_4 \vert \phi_c -\phi_\infty \vert \ge 2 \beta \vert \phi_N(R) \vert
\end{equation}
where $\phi_N (R) = -G_N \Delta M/R$ is Newton's potential at the surface of the overdensity. We will call this the {\it no-shell condition}.

The constant $D$ is determined from the boundary conditions to be $D = \beta \kappa_4 (\Delta M)/(4\pi)$ where $\Delta M = 4 \pi (\Delta \rho) R^3/3$.
Thus from Eq.~(\ref{outside}), the potential outside reads
\be
\kappa_4 \phi^+= \kappa_4 \phi_\infty - 2\beta \frac{G_N \Delta M}{r} \,.
\ee
As a result, from Eq.~(\ref{gravp}) the contribution of the chameleon to the force acting on tests particles is given by
\be
\Psi= \beta \kappa_4 \phi_\infty  -(1+2\beta^2) \frac{G_N \Delta M}{r}
\ee
The first term has no gradient, the second term gives the force on tests particles. It only involves the overdensity and its mass $\Delta M$. This corresponds to an increase of Newton's force by $(1+2\beta^2)$
\begin{equation}
\gamma -1=2\beta^2
\end{equation}
as
obtained earlier for a test mass.

When discussing the gravitational collapse of an initial overdensity, if the density perturbation is small and the size of the overdense region is not too large, then the inside of the collapsing sphere will be described by the results in this section. This is true as long as $\phi_b$ does not converge towards $\phi_c$, i.e. when the no-shell condition is satisfied. When the no-shell condition is violated, the behaviour of the chameleon profile changes drastically.

\subsection{Large bodies, thin shells}
For large bodies, we expect the value of the field at the centre of the spherical body to be $\phi_c$, i.e.
when Newton's potential is large enough, the solution is identically constant inside a shell
\be
\phi^-(r) =\phi_c \qquad (0\le r \le R_s)
\label{inside}
\ee
(see Eq.~(\ref{sss}) for $\phi_b=\phi_c$).  Following \cite{Khoury:2003rn},
in the shell
\be
\phi^s(r) =\phi_c+ \frac{\beta\kappa_4 \rho_c}{3} \left(\frac{r^2}{2} + \frac{R_s^3}{r}\right) -\frac{\beta \kappa_4 \rho_c R_s^2}{2} \qquad (R_s\le r\le R)
\ee
while outside the body
\be
\phi^+(r)=\phi_\infty - \frac{\beta \kappa_4 \rho_c (R^3-R_s^3)}{3(1+m_\infty R)}  \frac{e^{-m_\infty (r-R)}}{ r} \qquad (r\geq R).
\ee
where we have not assumed anything about the magnitude of $m_\infty R$.
Notice that the field outside the body is only determined by the mass inside the shell.
The radius of the shell itself, $R_s$ is determined from continuity of the fields at $R$, namely $\phi^s(R)=\phi^+(R)$, or
\be
\phi_c-\phi_\infty= \frac{1}{3}\kappa_4 \beta \rho_c \frac{R_s^3}{R}(1-\frac{1}{1+m_\infty R}) +\frac{1}{6}\kappa_4 \beta \rho_c R^2 (1+\frac{2}{1+m_\infty R})-\frac{1}{2} \kappa_4 \beta \rho_c R^2_s
\ee
When $m_\infty R\ll 1$, this reduces to
\be
\phi_c -\phi_\infty = \frac{ \kappa_4 \beta \rho_c}{2} (R_s^2 -R^2).
\label{detshell}
\ee
In the following we will always assume that $m_\infty R\ll 1$ which will correspond to the astrophysical situation of interest.

Let
\be
R_s = R-\Delta R
\ee
Then to first order in $\Delta R/R$, Eq.~(\ref{detshell}) reduces to
\be
\frac{\Delta R}{R} =  \frac{\phi_\infty - \phi_c}{6 \beta \vert \Phi_N\vert  M_{\rm Pl}}
\label{japan}
\ee
where $\Phi_N(R) = -G_N M/R$ is the total Newtonian potential at the surface of the body (see Eq.~(\ref{totalPhi})) and $M=4 \pi R^3 \rho_c/3$ is the mass of the body.
Eventually when the radius of the compact body is large enough, the shell appears.
More precisely, this occurs at a radius $R_c$ which, from Eq.~(\ref{japan}), is determined by
\be
\kappa_4 (\phi_\infty -\phi_c)= 6 \beta  \vert \Phi_N(R_c)\vert.
\label{japan2}
\ee
When $ (\phi_\infty -\phi_c)\ll 6 \beta M_{\rm Pl}\Phi_N(R_c)$, the shell is really thin.

The modification of Newton's law outside the compact body is given by
\be
\Psi=\kappa_4 \beta \phi_\infty- \frac{G_N \Delta M}{r} - 2\beta^2 \left(1- \frac{R_s^3}{R^3}\right)\frac{G_N M}{r}.
\label{japan3}
\ee
Notice that the modification of gravity is entirely due to the mass inside the shell. The rest of the mass of the spherical body is completely shielded. The Laplacian of Newton's potential $\Psi$ at the surface of the
overdensity is
\begin{equation}
\left. \Delta  \Psi \right|_{r=R} ={4\pi G_N} (\rho_c - \rho_\infty)  +{8 \pi G_N}\beta^2 \left( 1- \frac{R_s^3}{R^3} \right) {\rho}_c
\label{Ox}
\end{equation}
whilst the total Newtonian potential $\Psi_N$ including the contribution of the cosmological density inside the sphere, see Eq.~(\ref{geo}), is given by
\begin{equation}
\left. \Delta  \Psi_N \right|_{r=R} ={4\pi G_N}\rho_c  +{8 \pi G_N}\beta^2 \left( 1- \frac{R_s^3}{R^3} \right) {\rho}_c .
\end{equation}
This is the Newtonian potential multiplied by
\begin{equation}
\gamma(R,R_s)= 1 + 2\beta^2 \left(1- \frac{R_s^3}{R^3} \right)
\label{gammaRRs}
\end{equation}
extrapolating between General Relativity ($\gamma=1,\ R_s=R$) and a small body regime ($\gamma=1+2\beta^2, \ R_s=0$).

\section{Spherical Collapse}

In the previous section we have found a description of the modification of gravity for chameleon theories in case of a static, spherically symmetric, overdensity. We now generalise this discussion to the dynamical situation in which the density inside and outside the body evolve on cosmological time scales.

\subsection{Initial conditions}

Before writing down the set of nonlinear equations governing the evolution of the system, it is very useful to analyse the astrophysical situation in which an initial inhomogeneity appears as a spherical body of
radius $R$ immersed in a homogeneous background.
Depending on the radius, the initial body may or may not have a chameleonic shell.

The radius $R_c$ below which no shell is present is given implicitly in Eq.~(\ref{japan2}).
This simplifies when $ \rho_c= (1+\delta_i) \rho_\infty$ with $\delta_i\ll 1$.
First of all the minimum of the effective potential is such
that
\be
\kappa_4(\phi_c-\phi_\infty)= -\frac{\beta \kappa_4^2 \rho_\infty }{m_\infty^2} \delta_i
\ee
and therefore from Eq.~(\ref{japan2})
\be
R_c=  \frac{\sqrt \delta_i}{ m_\infty}.
\label{Rthin}
\ee
 For large objects with $R\ge R_c$, we will use the thin shell solution\footnote{We always check that the condition $m_\infty R \ll 1$ is verified.}. In particular, this will imply that the gravitational collapse of these objects is not too strongly modified compared to the General Relativity case.

\subsection{Collapse}
\label{ssec:collapse}

Hence let us consider an initial spherical overdensity with a thin-shell, thus the initial radius $R_i > R_c$ so that initially the thin shell is between $R_s \leq r \leq R_i$.  We now try to understand qualitatively the evolution of the system.  As we shall see, the dynamics of the nonrelativistic particles inside the overdensity depend crucially on whether they are within the thin shell or not.

Let $q^i$ denote the initial position of the particle. As shown in the previous section, see Eq.~(\ref{inside}), if $\vert q^i\vert \le R_s$, the chameleon field is constant and the particle experiences no extra force compared to gravity
\begin{equation}
\gamma (r)= 1, \qquad r\le R_s.
\end{equation}
So initially, the particles on a spherical shell of radius $\vert q^i\vert \le R_s$ collapse as in General Relativity. For particles on shells  within the thin-shell region itself, the initial force they feel is enhanced compared to gravity and
\begin{equation}
\gamma(r)= 1+ 2\beta^2 \left(1- \frac{R_s^3}{r^3} \right) \, ,\qquad r \geq R_s .
\end{equation}
This implies that the spherical collapse of these spherical shells $\vert q^i\vert \ge R_s$ is initially accelerated compared to the shells with  $\vert q^i\vert \le R_s$. As a result, due to the relative acceleration felt by particles at different radii, the density inside the thin-shell will not remain uniform: at a given instant $t$ and following the particles inside the shell in a Lagrangian way, the thin-shell density becomes
\begin{equation}
\rho (r,t)= \rho (q) {\rm det} (\frac {\partial q^i}{\partial r^j}).
\end{equation}
Of course this density depends on the dynamics of the nonrelativistic particles and therefore on the chameleon field in a bootstrap fashion.

The radius of the thin-shell $R_s$, defined as the boundary where no scalar force is felt,  starts moving too, thus $R_s = R_s(t)$.
For radii smaller than $R_s(t)$ we have
\begin{equation}
\phi=\phi_c,\qquad r\le R_s(t).
\end{equation}
Mass conservation implies that the mass initially outside the thin shell with $r\le R_s (t)$ is conserved
\begin{equation}
M_s= \frac{4\pi \rho_c(t)}{3} R_s^3(t).
\end{equation}
with a dynamical evolution as in General Relativity.
Inside the thin shell the chameleon field satisfies a Poisson equation as the potential contribution is negligible \cite{Khoury:2003rn}
\begin{equation}
\Delta \phi= \beta \kappa_4 \rho(r,t), \qquad R_s(t)\le r\le R(t)
\end{equation}
where both the inner radius $R_s(t)$ and outer radius $R(t)$ of the shell are time-dependent. The density inside the shell is not uniform now. Let us define the mass inside the shell
\begin{equation}
M(r,t)= 4\pi \int_{R_s(t)}^r dr' r'^2 \rho(r',t)
\end{equation}
Gauss' law implies  that
\begin{equation}
\phi(r,t)=\beta \kappa_4 \int_{R_s(t)}^r \frac{M(r',t)}{4\pi r'^2} dr' +\phi_c
\end{equation}
This allows one to evaluate the force due to chameleon on a shell of radius $r$ and therefore determine the motion of the nonrelativistic particles and the density of matter in the thin-shell.

Outside the body, Gauss' law allows one to write the outer solution as
\begin{equation}
\phi= \phi_\infty - \beta \kappa_4 \frac{M(R(t))}{4\pi r}
\end{equation}
as long as $m_\infty r\ll 1$. Matching at the surface $r=R(t)$ the inner and outer solutions gives immediately
\begin{equation}
\phi_\infty -\phi_c = \beta \kappa_4 \int_{R_s(t)}^{R(t)} \frac{M(r',t)}{4\pi r'^2} dr' + \beta \kappa_4\frac{M(R(t))}{4\pi R(t)} \, .
\end{equation}
This equation relates the radius of the outer shell $R(t)$ to the radius of the inner shell $R_s(t)$ and $\rho(r,t)$. Notice that these equations are the generalisation of the ones obtained in section 3.2 when the density inside the spherical body is both constant in time and spatially uniform. Here this is not the case anymore due to the collapse of the spherical body.

The previous description of the spherical collapse with the development of an inhomogeneous thin-shell due to the different accelerations felt by the nonrelativistic particles at different radii inside the thin shell is very difficult to solve numerically. It involves a boostrap approach as the chameleon field in the thin shell is determined by the matter density, which in turn is determined by the motion of the matter particles under the influence of the chameleonic field. In the following, we will not attempt to numerically solve this tough problem. We will work with the lowest approximation assuming that the density inside the shell remains uniform spatially. More precisely we will follow the same approximation scheme as in the gravitational collapse with no chameleon. We assume that there is no shell-crossing and therefore the spherical shells at ordered radii $r_1<r_2$ remain ordered. This implies that the mass in each shell is constant in time. In this case, starting from a homogeneous distribution of matter we have
\begin{equation}
M(r,t)= M \frac{r^3-R_s^3(t)}{R_i^3}
\end{equation}
where $R_i$ is the size of the initial overdensity and $M$ its total mass. As a result we find that
\begin{equation}
\phi(r,t)=\beta\kappa_4 \frac{M}{4\pi R_i^3}\left( \frac{r^2}{2} + \frac{R_s^3(t)}{r} -\frac{3}{2} R_s^2(t) \right) +\phi_c
\end{equation}
This corresponds to the solution with a uniform  density $\rho_c$ in the shell as already obtained in section 3.2.
This first order approximation will give  us a reliable description of the collapse when the size of the thin shell is very small and inhomogeneities have little time to develop. Of course, a more thorough study of the full set of equations is required to assess the full validity of this approximation. In general, shells will cross in a finite time leading to the formation of caustics. This is left for future work.

Moreover, when applying the results on the spherical collapse with a thin shell, one should be wary of the fact that in realistic cases for astrophysical overdensities, the presence of inhomogeneities breaking an exact spherical symmetry will lead
to even greater difficulties. Indeed, the modification of gravity induced by the chameleon will be nonspherical implying that matter particles will feel a nonspherical chameleon force which could therefore amplify the initial inhomogeneity. A full numerical simulation of the dynamics of a thin-shelled and a slightly inhomogeneous
overdensity is certainly worth carrying out.

\subsection{Sphere dynamics with a thin shell}

Assuming that the density inside the thin-shell remains uniform, we can write down a simplified equation for the dynamics of the spherical collapse.
Using the Poisson equation modified by the interaction of matter with the chameleon
field, Eq.~(\ref{geo}), and mass conservation one gets
\be
\frac{\ddot R}{R} = -\frac{4 \pi G_N}{3} \left[ \rho_\phi (1+3 w_\phi) +  \gamma(R,R_s) \rho_c \right]
\label{ddotR}
\ee
with $ \dot{} =  d/dt$ and $\gamma$ is given in Eq.~(\ref{gammaRRs}). This equation
includes the chameleon enhancement of gravity
taking into account the thin-shell effect, where $R_s$ is the
radius of the shell. The normal gravity case is obtained either for $\beta = 0$ (no coupling) or if
the shell is very thin, $R_s/R \rightarrow 1$.

We will assume that $\rho_\phi$ is constant ($w_\phi=-1$) and write
\be
\rho_c = \frac{3 M}{4 \pi R^3} = \frac{R_i^3 \bar{\rho}_i (1+\delta_i)}{ R^3} =
\frac{\Omega_m^{(0)} \rho_{cr}^{(0)} (1+\delta_i)}{\tilde{R}^3},
\ee
where again we have used mass conservation, and the usual assumption $\Omega \approx 1$:
\be
M = M_i = \frac{4 \pi}{3} R_i^3 \bar{\rho}_i (1+\delta_i),
\ee
$\bar{\rho}_i =\rho^i_\infty= \bar{\rho}_0 a_i^{-3}$ is the initial background density. We define $\tilde{R}(t)$ as the radius of the
collapsing spherical
region in units of the initial comoving radius $R_i/a_i$:
\be
\tilde{R}(t)= \frac{R(t)}{R_i/a_i}
\ee
with $\tilde{R}(t_i) = a_i$.
(Notice that, as explained above, we have assumed $\Omega \approx 1$, so that we can identify the energy densities in the Jordan and Einstein frames, and $\tilde{a}$ with $a$.)

Using
\be
\rho_{cr}^{(0)} = \frac{3 H_0^2}{8 \pi G_N}
\ee
we can write Eq.(\ref{ddotR}) in terms of $\tilde{R}$ as:
\be
\frac{\ddot{\tilde{R}}}{\tilde{R}} = -\frac{H_0^2}{2} \left(-2 \Omega^{(0)}_\phi + \frac{1}{\tilde{R}^3} \gamma(\tilde{R},\tilde R_s) \Omega^{(0)}_m (1+\delta_i) \right)
\label{ddotR2}
\ee
where we will see below that the only explicit dependence on $R_i$ arises in the
$\gamma$ factor.
\subsection{The linear regime}

We have seen that the chameleon dynamics and the existence of a thin shell modifies the collapse of an initial spherical overdensity.
It is particularly interesting to analyse the evolution of such a configuration in the linear regime extrapolated to values of the density contrast where the linear approximation is not valid anymore. Indeed,
within this approximation,
we can also compute the value of the  critical density, $\delta_c(z)$,
used in the Press-Schechter formalism for structure formation \cite{PS}.
It is defined as the linearly extrapolated density contrast from an initial value such
that collapse occurs at a given redshift $z$. Practically,
one must  first find the initial density contrast required for collapse at $z=0$.
Then  this initial density contrast evolves using the linearized form of
eq.(\ref{nonlineardelta}).

Let us first consider the linear evolution of an overdensity with a thin shell.
The Laplacian of Newton's potential $\Psi$ at the surface of the overdensity is given in Eq.~(\ref{Ox}):
\begin{equation}
\Delta  \Psi\vert_{r=R(t)}={4\pi G_N} \rho_\infty \delta  +{8 \pi G_N}\beta^2 \left( 1- \frac{R_s^3}{R^3} \right) {\rho}_\infty (1+\delta)
\end{equation}
In physical time $t$ and from Eq.~(\ref{nonlineardelta}), the linear equation for $\delta$ valid only in the thin shell regime is:
\be
\ddot{\delta} + 2 H \dot{\delta} - 4 \pi G_N \left[1+ 4 \beta^2 \left( 1- \frac{R_s^3}{R^3} \right) \right] {\rho}_\infty \delta = 4 \pi G_N \left[2 \beta^2 \left( 1- \frac{R_s^3}{R^3} \right) \right] {\rho}_\infty \, .
\ee
For $\beta =0$ (no coupling) or $R_s = R$ (complete shielding) one recovers the standard
linear evolution equation. Let us analyse briefly the structure of this equation. First of all there is a forcing term
which arises since the chameleon couples to all matter inside the sphere, whereas
only the perturbed Newtonian potential contributes to the growth of perturbations.
Then the term linear in $\delta$ receives two contributions in $2\beta^2$ when $R_s \to 0$. This follows from the $2\beta^2 (1+\delta)^2$ term in the full nonlinear equation.
This equation is valid as long as the thin shell condition is valid, i.e. as long as $R_s \ge 0$. In particular, the thin shell is very thin when $ \kappa_4\vert \phi_c-\phi_\infty \vert \ll 6\beta \vert \Phi_N(R(t))\vert$ where $\Phi_N (R(t))$ is total Newton potential of the sphere. When the sphere is small enough and/or sparse enough, the thin shell regime is no longer valid. In this case, the field configuration for the chameleon is different. In particular we have seen that as soon as $ \kappa_4 \vert \phi_c -\phi_\infty\vert \ge 2 \beta \vert \phi_N(R(t))\vert$, where $\phi_N(R(t))$ is the Newton potential due to the overdensity only, the value of the chameleon at the centre of the sphere is not $\phi_c$ anymore. In this regime, the Laplacian of $\Psi$ is simply:
\begin{equation}
\Delta  \Psi\vert_{r=R(t)}= 4\pi G_N \rho_\infty \delta  +8 \pi G_N\beta^2  {\rho}_\infty \delta
\end{equation}
where the first term comes from $\phi_N$ and the second from the chameleon. In this case, the linear equation reduces to:
\be
\ddot{\delta} + 2 H \dot{\delta} - 4 \pi G_N \left[1+ 2 \beta^2  \right] {\rho}_\infty \delta = 0
\ee
Notice that the forcing term has disappeared and the effective Newton constant is simply $(1+2\beta^2)$ larger than $G_N$. This is nothing but the usual linear equation deep inside the Compton wavelength of the chameleon field. This is justified here as this equation is valid for radii $R(t)\ll \sqrt{\delta}/m_{\infty}$.

There is a sharp difference between the two regimes. When no shell is present, the linear evolution equation does not depend on $R$ and therefore the critical density contrast does not depend on the initial value $R_i$. This is not the case anymore in the thin shell regime. Indeed, the evolution equation is sensitive to the ratio
$R_s/R$ which depends on the initial size $R_i$. As a result, the critical density contrast becomes scale dependent. This type of behaviour leads to a moving barrier problem for structure  formation. The study of the connection between moving barriers and chameleons is left for future work.

\section{Phenomenology}

\subsection{Inverse power law models}

In the following we will focus on models defined by an inverse power law potential and a constant coupling to matter~\cite{Khoury:2003rn,braxkhoury}. These models are the original chameleon ones and are defined by the potential
\be
V(\phi)= \Lambda_0^4 + \frac{\Lambda^{4+n}}{\phi^n}+ \dots
\label{powerlaw}
\ee
where we neglect higher inverse powers of the chameleon field.
The effective potential has a minimum at
\be
\phi_{min}=\left(\frac{n M_{\rm Pl} \Lambda ^{4+n}}{\beta
\rho} \right)^{1/(n+1)}.
\ee
and the mass of the field at the minimum is given in Eq.~(\ref{msqdef}).
For this model (assuming $\phi \ll M_{\rm Pl}$),
\be
m^2= (n+1) \beta \left( \frac{\rho}{M_{\rm Pl}^2} \right) \left( \frac{M_{\rm Pl}}{\phi_{min}}\right) \gg (n+1) \beta \left( \frac{\rho}{M_{\rm Pl}^2} \right).
\label{mmin}
\ee
In a cosmological setting, the Hubble parameter $H \simeq \sqrt{\rho}/M_{\rm Pl}$ so that $m\gg \sqrt{\beta} H$.
Hence, the field quickly relaxes to the minimum of the potential and
sits there for most of the cosmological history, at least since matter equality.
The fact that the scalar field can have a mass much larger than $H$ is one of the main differences between
chameleon models and usual quintessence models.
At the minimum of the potential,
\be
V_{eff}(\phi_{min}) \approx \Lambda_0^4 + \rho.
\ee
This implies that the model behaves like a pure cosmological constant in the recent past. The matter contribution is simply the usual energy density entering the Friedmann equation.  Hence the main difference between these chameleon models and a $\Lambda$CDM model will be at the level of the growth of structures only.

Alternatively, one can also  characterise the chameleon potentials by the evolution of the chameleon mass  $m_\infty(a)$, and the coupling $\beta$ with the Cold Dark Matter sector.
In the appendix we show that a power law evolution of the mass $m_\infty(a) = m_i (a/a_i)^r$ with $n=-(6+2r)/(3+2r)$ leads to a power-law potential
such as Eq.~(\ref{powerlaw}) and we explicitly compute $\Lambda$ as a function of $m_i$ and $r$ for this case. In the following we will utilise the inverse power law as a useful parameterisation of the chameleon potential valid on cosmological scales starting from  matter equality,  when the matter density evolves from $\rho\approx 10^{-17}{\rm g/cm^3}$ to $\rho\approx 10^{-29}{\rm g/cm^3}$.

In order for the chameleon effects to be significant, the  Compton length scale $m_\infty^{-1}$ in the cosmological background should be larger than the
scale of the perturbation $R$. As we will study perturbations at the Mpc scale, we require the initial mass $m_i$ at matter equality to be $m_i^{-1} = {\cal O} (0.02\mbox{Mpc})$. Since we typically have small initial perturbations
of the order of $\delta_i = 10^{-4}$ at $z_i=10^{4}$,  we estimate $R_c = {\cal O}(10^{-4} \mbox{Mpc})$
from Eq.~(\ref{Rthin}) and we
are justified to use the  shell solutions for  scales larger than $R_c$. In terms of comoving scales, we consider scales ranging from
$R_i/a_i=1$ to $100$~Mpc for which the condition
$m_\infty R\ll 1$ is satisfied. Moreover, a thin shell is initially present for these scales.

%In particular,
%in these models,
%we can estimate $R_c$ defined in Eq.~(\ref{Rthin}). Using (\ref{mmin}) and $\rho_i = \rho_0 a_i^{-3}$
%one gets
%\be
%R_c=  \frac{\sqrt \delta_i}{m_\infty} \simeq  \sqrt{\delta_i}
% \left[ \frac{ n^{1/(n+1)}}{n+1} \beta^{-\frac{n+2}{n+1}} a_i^{\frac{3(n+2)}{n+1}}\right]^{1/2}
%\left[ \frac{M_{\rm Pl} \Lambda^3}{\rho_0} \right]^{\frac{n+2}{2 (n+1)}} \frac{1}{\Lambda}
%\ee
%Thus, for example, for  $\rho_0 = 10^{-48}$ GeV$^4$, $\Lambda \sim 10^{-12}$ GeV, $M_{\rm Pl} \sim 10^{18}$ GeV, %$n=0.2$, $\beta=1$, $\delta_i=10^{-2}$  and $a_i=10^{-4}$ we obtain
%$R_c \sim 6 \times 10^{-13}$ Mpc. This justifies the use of the thin shell solution in our analysis as it is a %tiny scale and all scales of interest for the spherical collapse model will be much larger than this critical %radius.

When a shell develops, its radius depends on the nonlinear dynamics of the chameleon field and is determined by Eq.(\ref{detshell})
\be
\frac{R_s^2}{R^2}= 1- \frac{2}{\kappa_4 \beta \rho_c R(t)^2}(\phi_\infty -\phi_c)
\ee
where
\be
\phi_c= \left(\frac{4\pi R(t)^3 nM_{\rm Pl} \Lambda^{n+4}}{3\beta M}\right)^{1/(n+1)}
\ee
and similarly for $\phi_\infty$ with the background scale factor $a(t)$ replacing $R(t)$.

We can estimate the size of the thin shell by rewriting:
\be
\frac{\phi_c}{\Lambda}
= \left(\tilde{R}^3(t)\frac{n M_{\rm Pl} \Lambda^3}{3\beta \Omega_m^{(0)} \rho_{cr}^{(0)}(1+\delta_i)} \right)^{1/(n+1)},\ \ \frac{\phi_\infty }{\Lambda}= \left(a^3(t)\frac{n M_{\rm Pl} \Lambda^3}{3\beta \Omega_m^{(0)} \rho_{cr}^{(0)}} \right)^{1/(n+1)}.
\ee
Hence we obtain
\begin{eqnarray}
\frac{R_s^2}{R^2}
&=& 1- \frac{2}{\beta \Omega_m^{(0)}(1+\delta_i)} \left( \frac{M_{\rm Pl} \Lambda}{\rho_{cr}^{(0)} (R_i^2/a_i^2)} \right)
\left(\frac{n M_{\rm Pl} \Lambda^3}{3\beta \Omega_m^{(0)} \rho_{cr}^{(0)}} \right)^{1/(n+1)} \times \nonumber \\
&&\tilde{R}(t)\left(a^{\frac{3}{n+1}}(t) - \tilde{R}^{\frac{3}{n+1}}(t) (1+\delta_i)^{-\frac{1}{n+1}} \right).
\label{shell1}
\end{eqnarray}
As time evolves $a(t)$ grows faster than $\tilde{R}(t)$,
which eventually stops growing and collapses to zero.

We can estimate the size of the thin shell for a given model.
As an example, for $\beta=0.01$, $n=0.02$  and $\Lambda = 1.3 $ eV (which follows from fixing $m_i = 50$ Mpc$^{-1}$, see appendix),
we find approximately:
\be
\frac{R_s^2}{R^2} = 1- \left(\frac{1.6 \times 10^{20}\mbox{Mpc}}{R_i/a_i} \right)^2  \tilde{R}(t)\left(a^{\frac{3}{n+1}}(t) - \tilde{R}^{\frac{3}{n+1}}(t) \right)
\ee
Hence the initial thin shell disappears extremely quickly only to reappear when $\tilde R$ is very small, well after virialisation. In the case of inverse power law chameleons with large Compton wavelength $m_\infty^{-1}$, we find that the thin shell is only present for a very short part of the cosmological evolution since matter equality. This behaviour is typical of inverse power law chameleon models. Of course, it would be very interesting to study other classes of chameleon models such as the ones inspired by $f(R)$ models for which the thin shell may exist for a longer period of time.

\subsection{$f(R)$ models}

We can also apply our results to a class of $f(R) $ models defined by the action
\begin{equation}
S=\frac{1}{2\kappa_4^2}\int d^4 x \sqrt{-g} f(R)
\end{equation}
 These models are completely equivalent to the chameleon models we have presented\cite{braxshaw}.  In particular,  the coupling constant is uniquely fixed
\begin{equation}
\beta=\frac{1}{\sqrt 6}
\end{equation}
and the chameleon field is related to the curvature $R$ via
\begin{equation}
f'(R)=e^{-2\beta\phi/M_{\rm Pl}}
\end{equation}
while the potential is given by
\begin{equation}
V(\phi)=\frac{M_{\rm Pl}^2}{2} \left( \frac{ Rf'(R) -f}{f'^2(R)} \right)
\end{equation}
Inverse power law models can be retrieved using
\begin{equation}
f(R)= R- f_0 + \frac{p}{p+1} \bar R \left(\frac{R}{\bar R}\right)^{p+1}
\end{equation}
with $-1<p<0$ and $ \bar R \gg R\gg H_0^2$
leading to
\begin{equation}
V(\phi)= \Lambda_0^4 + \frac{M_{\rm Pl}^2}{2} \bar R\, \frac{p^2}{p+1} \left(\frac{-2\beta \phi}{pM_{\rm Pl}}\right)^{\frac{p+1}{p}}+\dots
\end{equation}
where $\Lambda_0^4=\frac{M^2_{\rm Pl} f_0}{2}$. Notice that $n=-\frac{p+1}{p}$.
In these models, as $\beta=1/\sqrt{6}$ the Compton wave-length $m_\infty^{-1}$ tends to be larger than the one of the example presented in the previous section. Of course, a more thorough study would be necessary to determine how the thin shell regime could appear or disappear with more general $f(R)$ models.

\subsection{The nonlinear regime}

\begin{figure}[h,b,t]
\begin{center}
			\includegraphics[scale=1.0]{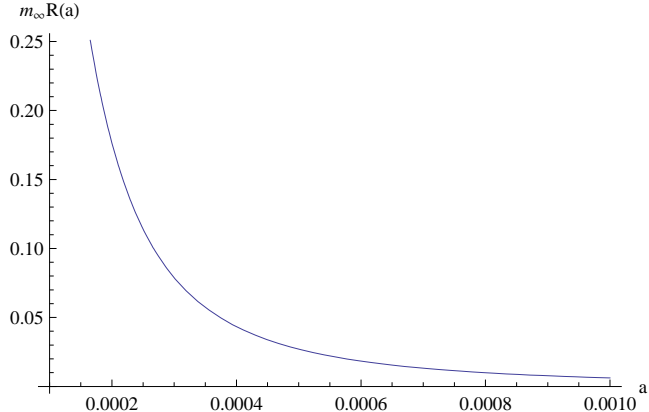}
\end{center}
\caption{Plots of $m_\infty R(a)$ as a function of the scale factor $a$ for $R_i/a_i = 100$ Mpc with initial
density contrast fixed such as to give collapse today for the usual gravity case. Other
parameters are given in the text.}
\label{Mass}
\end{figure}

We will now study the effects of the chameleon field on the evolution of spherical perturbations assuming that the inhomogeneities in the thin shell can be neglected.
We first solve numerically Friedmann's equation
for the background scale factor $a(t)$, with time in units of Hubble time $H_0^{-1}$.
We start the evolution of perturbations at $a_i = 10^{-4}$. In the examples that follow
we choose $\Omega_m^{(0)} = 0.25$, $\Omega_\phi^{(0)} = 0.75$, $\beta=0.01$, $n=0.02$ and $\Lambda = 1.3$ eV.
We numerically solve Eq.~(\ref{ddotR2}) for the nonlinear evolution of the size of the inhomogeneity.
We show in fig.~\ref{Mass} that the approximation $m_\infty R(a)<< 1 $
is valid for our example.

\begin{figure}[h,b,t]
\begin{center}
			\includegraphics[scale=1.0]{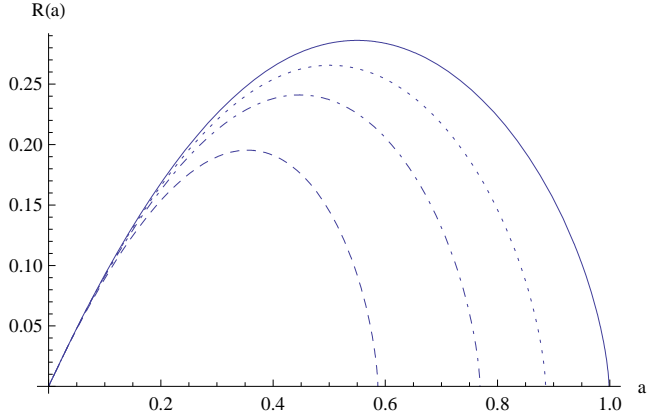}
\end{center}
\caption{Plots of $R(a)$ as a function of the scale factor $a$ for $R_i/a_i = 1,10$ (dashed and dot-dashed lines) and
$100$ Mpc (dotted line) compared with the behaviour for $R$ in usual gravity (solid line), with initial
density contrast fixed such as to give collapse today for the usual gravity case. Other
parameters are given in the text.}
\label{Rcollapse}
\end{figure}

The effect of the chameleon on the evolution of the radius of the spherical perturbation is shown
in fig.~\ref{Rcollapse}, where one sees that it
enhances the gravitational collapse, as expected.  In this example, we choose the same parameters
as above and $\delta_i = 10^{-3.428}$, for which the collapse occurs today
for the usual gravity case ($\beta=0$).

Whereas in the usual gravity case ($\beta=0$) the collapse time (or $a_{\rm collapse}$) is independent of $R_i$, this is no longer the case in the chameleon model.  Here, not only is the collapse time $R_i$ dependent, but the {\it larger} the initial radius, the {\it slower} the collapse.  This suggests that for initially large bodies, the outer shells lag behind the inner ones, never catching them up during collapse.
Following the discussion of section \ref{ssec:collapse}, we therefore expect that the density of these
objects will not remain uniform. Of course, the description that we have given is only applicable when the density is uniform in the thin shell. To gauge the validity of this approximation, we have tabulated in Table \ref{table1}
the scale factors at collapse for various initial radii. We see that large enough bodies collapse with a deviation which is around 10 percent. Even for smaller spherical overdensities like 10 Mpc, the difference is only of order 25 percent. We therefore expect that our results  are within the right ball-park.
\begin{center}
\begin{table}[h,b,t]
\hskip 4.5 cm
\begin{tabular}{|c|c|c|}
\hline
\hline $R_i/a_i ({\rm Mpc})$ & $a_{\rm collapse}$  \\
\hline LCDM & 1\\
\hline 1 &0.528 \\
\hline 10 & 0.769 \\
\hline 25 & 0.823 \\
\hline 50 & 0.857 \\
\hline 100 &  0.886\\
\hline
\end{tabular}
\caption{Values of the scale factor at collapse for different $R_i/a_i$.}
\label{table1}
\end{table}
\end{center}

\begin{center}
\begin{table}[h,b,t]
\hskip 4.5 cm
\begin{tabular}{|c|c|c|}
\hline
\hline $R_i/a_i ({\rm Mpc})$ & $\ln \delta_i$  \\
\hline LCDM & -3.428\\
\hline 1 & -3.613\\
\hline 10 &-3.497 \\
\hline 25 & -3.476 \\
\hline 50 & -3.464\\
\hline 100 & -3.455\\
\hline
\end{tabular}
\caption{Values of the scale factor at collapse for different $R_i/a_i$.}
\label{table2}
\end{table}
\end{center}

\begin{figure}[h,b,t]
\begin{center}
			\includegraphics[scale=1.]{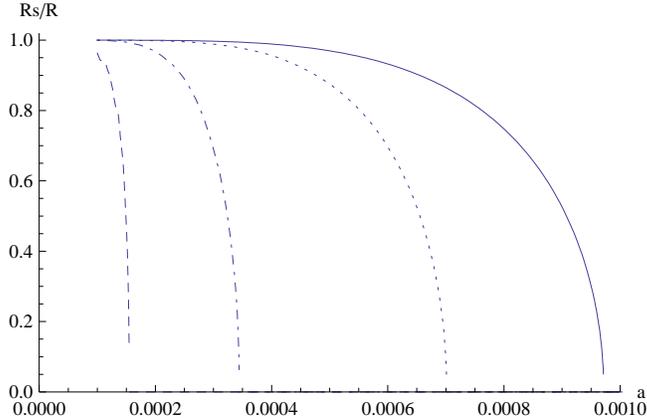}
\end{center}
\caption{The ratio $R_s/R$ as a function of the scale factor $a$ for $R_i/a_i=1, 10, 50 $ and $100$ Mpc (left to right) and the parameters given in the text.}
\label{shell}
\end{figure}

\begin{figure}[h,b,t]
\begin{center}
			\includegraphics[scale=1.]{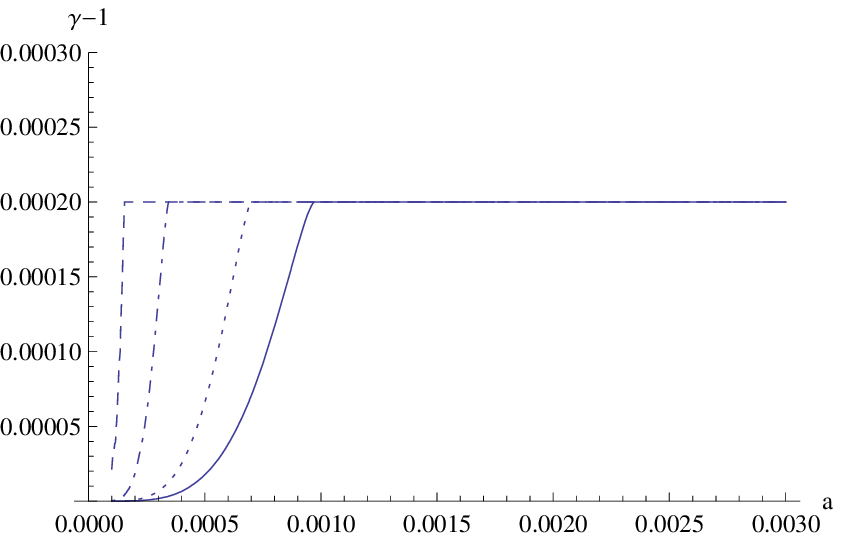}
\end{center}
\caption{The effect of modified gravity as measured by $\gamma-1$ as a function of the scale factor $a$ and for $R_i/a_i=1, 10, 50$ and $100$ Mpc  (left to right). }
\label{gamma}
\end{figure}

The thin shells can be seen in fig.~\ref{shell}, where we plot
$R_s/R$ for $R_i/a_i=1, 10, 50$ and $100$ Mpc, though here we have set $\delta_i$ such that collapse occurs today in each case. The shell disappears quickly here and the subsequent evolution corresponds to the no shell regime. The thin shell would reappear just before collapse well after virialisation. This effect can also be seen in fig.~\ref{gamma}, where we plot $\gamma -1$ which rapidly approaches the value
$\gamma-1 = 2 \beta^2 = 2 \times 10^{-4}$ in our numerical example.
The effect of  modified gravity can be seen by in the scale dependent values of the initial density contrasts for collapse today as in table \ref{table2}.

Using the thin shell linear equation, let us describe the behaviour of the critical density.
For standard gravity we find $\delta_c(0) = 1.672$, which is smaller than the usual
Einstein-de Sitter value of $\delta_c(0) = 1.686$, since the linear growth function is suppressed in this case.
For the chameleon case, one needs a smaller initial perturbation to reach collapse
today. However, the linear perturbation grows faster than in standard gravity and indeed we find a slightly larger values
of $\delta_c$ corresponding to the small value of $\beta$. For instance when $R_i/a_i=1$ Mpc, we obtain $\delta_c=1.677$. We expect that the discrepancy would be stronger for models where a larger value of $\beta$ would be  compatible with a large Compton wavelength $m_{\infty}^{-1}$.

We must emphasize
that in the chameleon case, the value of the critical density is {\it scale dependent}, leading
to a moving barrier problem for structure formation. Here this scale dependence is suppressed by the smallness of $\beta$ which is numerically necessary to reach large values of the Compton wave length. We expect that more general models of the $f(R)$ type could lead to a stronger scale dependence of the critical density contrast.
A fully detailed phenomenological analysis is left for further work.

\section{Conclusions}

We have investigated the effects of a chameleon scalar field on the nonlinear stages
of the gravitational collapse of a spherical overdensity region immersed in a FRW background.
The gravitational potential is effectively changed by the presence of the new long range
interaction. The modification depends on the profile of the scalar field. We studied the profile
for the situation at hand, finding two very distinct classes of profiles depending on the size of
the collapsing region. In particular, for large enough collapsing region, the nonlinear dynamics
of the scalar field produces what is  known as a thin shell effect, where the modification of gravity
is essentially shielded. This has an influence on the dynamics of the shell as well as a potential effect on the critical density contrast for collapse,
which in this regime of chameleon theories becomes scale dependent.

We have also studied the spherical collapse in the case of the original chameleon model with an inverse power law potential and a constant coupling to Cold Dark Matter.
The effects are larger for the smaller collapsing regions. In all cases, the initial thin shell disappears quickly and the collapse proceeds with no shell until
a potential thin shell reappears after virialisation. More general models such as $f(R)$ theories may lead to a stronger influence of the thin shell regime on the spherical collapse.

We have emphasized that an inhomogeneous matter density in the thin shell will certainly result from the influence of the chameleon of the spherical dynamics. The phenomenological results in this paper have been obtained neglecting this effect. Although we expect that this phenomenon will be negligible for large enough initial radii, for smaller radii we  expect that singularities such as caustics may appear.
A more detailed analysis of these effects is left for future work.

\section*{Acknowledgements}
We would like to thank Raul Abramo for his early collaboration
in this project and Patrick Valegas for useful comments. We also thank Douglas Shaw for pointing out the role of inhomogeneities in the thin shell and Marcos Lima for spotting a numerical inconsistency in a previous version.
This work was partially supported by a CAPES-COFECUB project (PhB and RR),
by a Fapesp Tematico project (RR) and a CNPq research fellowship (RR).
One of us (PhB) would like to thank the EU Marie Curie Research \& Training network ``UniverseNet" (MRTN-CT-2006-035863) for support and
DAS would also like to thank the Agence Nationale de la Recherche grant "STR-COSMO" (ANR-09-BLAN-0157) for support.
As we were completing our final version, \cite{wi}  appeared with some overlap with our paper.

\section*{Appendix: Reconstructing the Chameleon Potential}

Chameleons theory in the cosmological regime are determined by the coupling $\beta$ and the time dependent mass $m_\infty(a)$ as a function of the scale factor
as long as $m_\infty \gg H$ and the minimum of the effective potential is a dynamical attractor. Indeed, we can obtain
\begin{equation}
V''\equiv \frac{d^2V}{d\phi^2}= m^2_\infty (a) - \beta^2 \frac{\rho(a)}{M_{\rm Pl}^2}
\end{equation}
Using the minimum equation we deduce
\begin{equation}
\frac{d\phi}{dt}=-\frac{\beta}{M_{\rm Pl}} \frac{\dot \rho}{V''}
\end{equation}
implying that
\begin{equation}
\phi(a)= -\frac{\beta}{M_{\rm Pl}}\int_{a_i}^a \frac{d\rho}{da}\frac{1}{V''(a)} da +\phi_i \, .
\end{equation}
The minimum equation implies that
\begin{equation}
V=V_i +\frac{\beta^2}{M_{\rm Pl}^2} \int_{a_i}^a \rho\frac{d\rho}{da} \frac{1}{V''(a)} da \, .
\end{equation}
This defines the potential parametrically. Of course, one should check that local tests of gravity are then satisfied and that the long time behaviour of
the model is the one of a $\Lambda$-CDM model.

Let us give an explicit example:
\begin{equation}
m_\infty(a)= m_i(\frac{a}{a_i})^r
\end{equation}
In the matter dominated era, we have
\begin{equation}
V''(a)= m_i^2(\frac{a}{a_i})^{2r}- 3\beta^2 H_i^2(\frac{a_i}{a})^3
\end{equation}
Assuming that $m_i\gg H_i$, we can neglect the second term initially. Using $H(a)= H_i (\frac{a_i}{a})^{3/2}$, we obtain
\begin{equation}
\phi (a)= \phi_i + \frac{3\beta \rho_i}{m_i^2 M_{\rm Pl}} \frac{1}{3+2r}\left ( 1- (\frac{a_i}{a})^{3+2r}\right )
\end{equation}
As long as $3+2r<0$, we can choose $\phi_i$ such that
\begin{equation}
\phi= \frac{3\beta \rho_i}{m_i^2 M_{\rm Pl}} \frac{1}{\vert 3+2r\vert} (\frac{a_i}{a})^{3+2r}
\end{equation}
This is a power law behaviour and $\kappa_4\phi \ll 1$ as long as $\beta=O(1)$.
Similarly we obtain
\begin{equation}
V= \Lambda_0^4 + \frac{3\beta^2 \rho_i^2}{m_i^2 M^2_{\rm Pl}}\frac{1}{6+2r} (\frac{a_i}{a})^{6+2r}
\end{equation}
This leads to the usual chameleon potential where we can identify
\begin{equation}
n=-\frac{6+2r}{3+2r}
\end{equation}
Similarly we have
\begin{equation}
\Lambda^{4+n}=\frac{3\beta^2 \rho_i^2}{m_i^2 M^2_{\rm Pl}}\frac{1}{6+2r}(\frac{3\beta \rho_i}{m_i^2 M_{\rm Pl}} \frac{1}{\vert 3+2r\vert})^n
\end{equation}
This is the potential we have used in the text.

%Local constraints are satisfied provided\cite{Khoury:2003rn}
%\begin{equation}
%\Lambda < (\frac{6^{n+1}}{n})^{1/(n+4)} \beta_{\rm baryon}^{(n+2)/(n+4)}10^{(15n-7)/(n+4)} 10^{-12} {\rm GeV}
%\end{equation}
%where the coupling to baryons may differ from the coupling $\beta_{\rm baryon}\ne \beta$ to Cold Dark Matter.

%Casimir experiments impose strong constraints on $\Lambda$.
%Defining
%\begin{equation}
%\Lambda_c= (\frac{\Lambda^{4+n}}{\Lambda_0^4})^{1/n}
%\end{equation}
%and
%\begin{equation}
%\bar\Lambda= \frac{\Lambda_0^2}{\Lambda_c}
%\end{equation}
%the Casimir pressure due to the chameleon  can be expressed as
%\begin{equation}
%P_\phi= \Lambda_0^4 K(n) (\bar \Lambda d)^{-2n/(n+2)}
%\end{equation}
%Using
%\begin{equation}
%\Lambda_0^4= 6.93~10^{-7} {\rm mPa}
%\end{equation}
%and $\Lambda_0^{-1}=8200~{\rm nm}$, the strongest constraints are $P_\phi(162 {\rm nm})<22.3 {\rm mPa}$ and %$P_\phi (400{\rm nm})< 0.69 {\rm mPa}$.
%For instance at $z=10^3$ with $\Lambda=7\cdot 10^{-11}$~GeV, $\beta=0.25$ and $n=0.02$ we find that $\beta_{\rm %baryon}>10^6$ while
%$1/m_i=0.02$~Mpc. This corresponds to a comoving length of 20 Mpc. In this case, the Casimir constraints are also %satisfied.

\typeout{--- No new page for bibliography ---}


\begin{thebibliography}{99}


\bibitem{lyth}
D.~ H.~ Lyth and A.~ Riotto, Phys.Rept.314:1-146,1999,
[arXiv: hep-ph/9807278].





\bibitem{Copeland:2006wr}
  E.~J.~Copeland, M.~Sami and S.~Tsujikawa,
  Int.\ J.\ Mod.\ Phys.\  D {\bf 15} (2006) 1753.
\bibitem{caldwell}
R.~ R. ~Caldwell and  M.~Kamionkowski,  Ann.~Rev.~Nucl.~Part.~Sci.~59: 397-429, 2009.

\bibitem{trodden}
A.~ Silvestri and  M.~ Trodden,
Rept.Prog.Phys.72:096901,2009.

\bibitem{Brax:2009ae} Ph.~Brax, [arXiv:0912.3610].


\bibitem{uzan}J.~P.~Uzan,  Rev.\ Mod.\ Phys.\ {\bf 75} (2003) 403.



\bibitem{DEInt}
L.~Amendola, G.~C.~Campos and R.~Rosenfeld,
Phys.\ Rev.\ D {\bf 75} (2007) 083506;
Z.~K.~Guo, N.~Ohta and S.~Tsujikawa,
Phys.\ Rev.\ D {\bf 76} (2007) 023508.
e-Print: astro-ph/0702015


\bibitem{Bertotti:2003rm}
  B.~Bertotti, L.~Iess and P.~Tortora,
  %``A test of general relativity using radio links with the Cassini
  %spacecraft,''
  Nature {\bf 425}, 374 (2003).
  %%CITATION = NATUA,425,374;%%


\bibitem{DP}
T.~Damour and A.~M.~Polyakov,
Nucl.\ Phys.\ {\bf B423} (1994) 532.

\bibitem{DGP}
G.~R.~Dvali, G.~Gabadadze and M.~Porrati,
Phys.\ Lett.\ {\bf B485} (2000) 208.

\bibitem{Khoury:2003rn}
  J.~Khoury and A.~Weltman,
Phys.\ Rev.\ Lett.\ {\bf 93} (2004) 171104;
  %``Chameleon Cosmology,''
  Phys.\ Rev.\  D {\bf 69} (2004) 044026.
%  [arXiv:astro-ph/0309411].
  %%CITATION = PHRVA,D69,044026;%%

\bibitem{braxkhoury}
Ph.~Brax, C.~  van de Bruck, A.~ C.~ Davis, J.~ Khoury, A.~ Weltman
Phys.Rev.D70:123518,2004.

\bibitem{BraxLinear}
Ph.~Brax, C.~van de Bruck, A.-C.~Davis and A.~M.~Green,
Phys.\ Lett.\ {\bf B633} (2006) 441.

\bibitem{SphColl}
J.~E.~Gunn and J.~R.~Gott III, Astrophys.\ J.\ {\bf 176}, 1 (1972).

\bibitem{MartinoStabenauSheth}
M.~C.~Martino, H.~F.~Stabenau and R.~K.~Sheth,
Phys.\ Rev.\  D {\bf 79} (2009) 084013.

\bibitem{SchmidtLimaOyaizuHuIII}
F.~Schmidt, M.~Lima, H.~Oyaizu and W.~Hu,
Phys.\ Rev.\  D {\bf 79} (2009) 083518.

\bibitem{SchmidtLimaHu}
F.~Schmidt, M.~Lima and W.~Hu,
Phys.\ Rev.\ D{\bf 81} (2010) 063005.

\bibitem{DEFluc}
An incomplete list of references includes:
D.~F.~Mota and C. van de Bruck, Astron.\ Astrophys.\ {\bf 421}, 71 (2004);
N.~J.~Nunes and D.~F.~Mota, Mon.\ Not.\ Roy.\ Astron.\ Soc.\ {\bf 368}, 751 (2006);
N.~J.~Nunes, A.~C.~da Silva and N.~Aghanim, Astron.\ Astroph.\ {\bf 450}, 899 (2006);
C.~Horellou and J.~Berge, Mon.\ Not.\ Roy.\ Astron.\ Soc.\ {\bf 360}, 1393 (2005);
M.~Manera and D.~F.~Mota, Mon.\ Not.\ Roy.\ Astron.\ Soc.\ {\bf 371}, 1373 (2006);
B.~M.~Schaffer and K.~Koyama, Mon.\ Not.\ Roy.\ Astron.\ Soc.\ {\bf 385}, 411 (2008);
L.~R.~Abramo, R.~C.~Batista, L.~Liberato and R.~Rosenfeld,
JCAP {\bf 0711}:012 (2007);
Phys.\ Rev.\  D {\bf 77} (2008) 067301;
Phys.\ Rev.\  D {\bf 79} (2009) 023516;
L.~R.~Abramo, R.~C.~Batista and R.~Rosenfeld,
JCAP {\bf 0907}:040 (2009);
S.~Basilakos, J.~C.~B.~Sanchez and L. Perivolaropoulos,
Phys.\ Rev.\  D {\bf 80} (2009) 043530;
F.~Pace, J.-C.~Waizmann and M.~Bartelmann,
{\tt arXiv:1005.0233 [astro-ph]};
P.~ Creminelli, G.~ D'Amico, J.~ Noreña, L.~ Senatore and F~ Vernizzi,
 {\tt arXiv:0911.2701};
P. ~Valageas,
Astron.Astrophys.508:93-106,2009.




\bibitem{dgg}
T.~Damour, G.~W.~Gibbons and C.~Gundlach,
Phys.\ Rev.\ Lett.\ {\bf 64} (1990) 123-126.


\bibitem{Babichev:2009fi}
  E.~Babichev and D.~Langlois,
  %``Relativistic stars in f(R) and scalar-tensor theories,''
  {\tt arXiv:0911.1297 [gr-qc]}.
  %%CITATION = ARXIV:0911.1297;%%

\bibitem{Damour1}
T.~Damour and G.~Esposito-Far\`ese, Phys.\ Rev.\ Lett.\ {\bf 70} (1993) 2220-2223.

\bibitem{Damour2}
T.~Damour and G.~Esposito-Far\`ese, Class.\ Quant\ Grav.\ {\bf 9} (1992) 2093-2176.

\bibitem{PS}
W.~H.~Press and P.~Schechter,
Astrophys.\ J.\ {\bf 187} (1974)425-438.

\bibitem{braxshaw}
Ph.~Brax, C.~ van de Bruck, A.~C.~ Davis and  D.~ J.~ Shaw,
Phys.Rev.D78:104021,2008.
\bibitem{wi}
N.~Wintergerst and V.~Pettorino,
{\tt arXiv:1005.1278 [astro-ph]}.


\end{thebibliography}
\end{document}